\documentclass[onecolumn,floatfix]{revtex4}
\usepackage{graphicx}
\usepackage{color}
\usepackage{array}
\usepackage{longtable}

\begin{document}
\title{Allometric Scaling in Scientific Fields}
\author{Hongguang Dong $^{1}$}
\author{Menghui Li $^{2}$\footnote{H. Dong and M. Li contribute equally}}
\author{Ru Liu $^{2}$}
\author{Chensheng Wu$^{2}$}
\author{Jinshan Wu$^{3}$\footnote{Corresponding author: jinshanw@bnu.edu.cn}}
\affiliation{
1.  Higher Education Press, Beijing, 10000, P. R. China\\
2. Beijing Institute of Science and Technology Intelligence, Beijing, 100044, P.R China \\
3. School of Systems Science, Beijing Normal University, Beijing, 100875, P.R. China
}

\begin{abstract}
Allometric scaling can reflect underlying mechanisms, dynamics and structures in complex systems; examples include typical scaling laws in biology, ecology and urban development. In this work, we study allometric scaling in scientific fields. By performing an analysis of the outputs/inputs of various scientific fields, including the numbers of publications, citations, and references, with respect to the number of authors, we find that in all fields that we have studied thus far, including physics, mathematics and economics, there are allometric scaling laws relating the outputs/inputs and the sizes of scientific fields. Furthermore, the exponents of the scaling relations have remained quite stable over the years. We also find that the deviations of individual subfields from the overall scaling laws are good indicators for ranking subfields independently of their sizes.

\end{abstract}

\maketitle

Some scientific fields might have many scientists and generate many publications while some might have small amount of researchers but with disproportionably larger publications. Have you wonder ever what is the relationship between the number of scientists and the number of papers (also number of references and received citations etc.) in scientific fields, and furthermore, whether and how such a relation can be used to indicate developmental stages of scientific fields? This question attracted considerable attention \citep{country1,country2,Innovation,Chinese}. In general, scaling laws are helpful to answer the above questions. For example, it is found that scaling laws are common phenomenon in scientific fields, including power law correlations between number of papers and of received citations \citep{country1,country2,Innovation} as well as between economic indicators and bibliometric measures \cite{Chinese}. In addition, a scale-independent indicator has been presented  to evaluate the research performance \cite{country2}. These studies help us to understand the performance of research units in terms of universities, cities and countries etc. In this study, we present a scaling analysis between size, which is measured by the number of authors, and input/output, where the former is represented by the number of references and the latter refers to number of papers and of received citations, of subfields at various levels. In a sense, we treat subfields as the universities, cities and countries in previous studies \cite{country1,country2,Innovation,Chinese}.

The formation, flourishing and decline of scientific fields, like the formation, rise and fall of cities, countries and industrial sectors \cite{Growth,interaction}, certainly comprise a question worth studying, although of a more abstract nature since the boundaries between scientific fields are less well defined than those of, e.g., cities.  Several researchers have begun to study the dynamic evolution of scientific fields \cite{evolution,modeling} or to evaluate the scientific performances of universities \cite{university1,university2}, research groups \cite{group1,group2,group3} and metropolitan areas \cite{Metropolitan}.

We note that in studies of cities, countries and many other systems, analyses of the scaling between their outputs/inputs and their sizes play an important and unique role. It has been found that patterns described by scaling laws commonly arise in complex systems; examples include typical scaling laws in biology, ecology and urban development \cite{allometry,theory}. These patterns facilitate the understanding of underlying mechanisms, dynamics and structures common to all cities and creatures \cite{theory}.  In biology, Kleiber's law describes the scaling relation between metabolism and body size \cite{Biology1,Biology2,Biology3}. The quarter-power exponents governing this relation originate from the physical and geometric properties of the underlying resources and information distribution network structures \cite{Biology5,Biology6}. In addition to metabolism, the relations between important biological variables (e.g., heart beat frequency, life span, and fertility rate) and body size also follow power laws \cite{Biology4}. In social systems, scaling relations are observed between urban indicators and city size \cite{Invention,Growth,Settlement,Supply,urban,road,countries}. Bettencourt et al. systematically studied the scaling relations between various properties of cities, including inventors, total wages, GDP, total housing, knowledge production and population \cite{theory,Invention}. All of these properties can be classified into three groups according to their scaling exponents \cite{theory,Invention,Growth,Settlement,Supply,urban,road,countries,Metropolitan}: economic, knowledge and innovation properties follow superlinear relations (increasing returns); household properties follow linear relations (individual human needs); and energy consumption and infrastructure properties follow sublinear relations (economies of scale) \cite{Growth}.  These results reveal a universal social dynamic underlying all these phenomena. For example, in the growth or stable stage of a city, it is important to have a superlinear relation between output and size and a linear or sublinear relation between input and size. One can imagine that if this is the case, then there must be some kind of mechanism to guarantee that such scaling laws are achieved. Therefore, scaling laws for the quantities of such systems imply the existence of such a mechanism and, consequently, the existence of essential interactions among the entities of the systems \cite{interaction}. Thus, such scaling laws require further investigation. In addition, scaling laws for cities provide a baseline for assessing the stages of development of specific cities independently of their population sizes. Therefore, an understanding of the scaling laws for cities can help policymakers to enhance the performance of their city relative to this baseline \cite{interaction}. In addition to studies treating cities as the targets of interest, studies have also been conducted on similar scaling relations between the outputs/inputs and sizes of metropolitan areas \cite{Metropolitan}, universities \cite{university1,university2} and countries \cite{country1,country2}. In principle, the values of the exponents in such scaling laws for various cities at various times might also be informative regarding, for example, whether more or less increasing returns are achieved with the scientific, technological and political development of civilization. However, such an analysis would require an enormous amount of historical data.

In this work, we perform an allometric scaling analysis of scientific fields. In other words, we treat scientific fields as the targets of interest, analogously to cities or biological/ecological systems, and study the scaling relations between the outputs/inputs and sizes of the investigated scientific fields. First, we have decades of reliable data that we can use to investigate the above question of co-evolution. Second, since scientific progress is less well known to the general public compared with the expansion or recession of cities, and also because science is more abstract than physical/financial measures of cities, we believe that such a study on scaling in scientific fields should be of particular interest. Compared with the degree of awareness regarding the driving forces and mechanisms of urban development, we believe that the driving forces and mechanisms of the development of scientific fields are even less well understood.

Similarly to the way in which the quantities related to individual cities are studied in allometric scaling analyses of cities, here, we study the quantities related to individual subfields of several disciplines, namely physics, mathematics and economics. In these three disciplines, a large number of papers have been classified into various subfields according to well-established subject classification schemes for each discipline. By using data for each paper, including its subject labels, authors and references as well as the other papers that cite it (its received citations), we first wish to examine whether such allometric scaling relations exist between the outputs/inputs and sizes of these scientific fields. Here, for a given subfield, we regard the number of authors who have produced papers in our datasets as its size, the number of references as its input, and number of papers or received citations as its output. Of course, the implied goal and logic are the same as those in studies of such relations for, e.g., cities, metropolitan areas, countries and universities: if allometric scaling relations exist, then mechanisms that give rise to such scaling relations also exist, which require further investigation. Second, if such scaling relations exist, then we also wish to infer the relative position or developmental stage of each subfield by examining the deviations of the subfields from the overall scaling relations and by investigating the evolution of the scaling exponents.

In this work, we show that the scientific organization and dynamics that relate the division of labor to scientific development and knowledge creation are very general and manifest as nontrivial quantitative patterns common to all subfields. We present an extensive body of empirical evidence showing that the outputs and inputs are scaling functions of subfield size that are quantitatively consistent across different disciplines and times. As shown later, we find a weak superlinear relation between output and size for these subfields. This indicates that the three disciplines in fact show similar productivity. In addition, we show that to a certain degree, the developmental stages of individual subfields as inferred from their deviations from the values expected according to scaling law are reasonable. We also find a weak superlinear relation between input and size. Furthermore, we find that during the few decades for which we have data, the values of the exponents have remained quite stable. Although it is commonly believed that science, as measured simply in terms of the number of papers or other indicators, has been developing much more rapidly in recent decades and possibly even in recent years, it seems that the underlying mechanism has remained the same, and consequently, the exponents relating output/input and size have also stayed the same.

\section{Data and method}

\paragraph{Datasets.} In our datasets, each paper is represented by a data entry that includes the year of publication, the subject classification code, and the numbers of author(s), reference(s) and citations as recorded in the Web of Science. We use the established subject classification scheme for each discipline to identify the subfields to which each paper belongs. The classification schemes used in this work are the Physics and Astronomy Classification Scheme (PACS) for physics, the Mathematics Subject Classification (MSC) for mathematics and the Journal of Economic Literature (JEL) codes for economics. These schemes are all hierarchical, and in this work, we use the fourth level of the physics classification scheme (e.g., 03.67), the third level of the mathematics classification scheme (e.g., 92B) and the second level of the economics classification scheme (e.g., N3).

The physics dataset is a collection of all papers published by the American Physical Society (APS) Physical Review journals from $1976$ to $2013$.  Here, we consider only those research papers, e.g., articles, brief reports and rapid communications, with PACS numbers. In total, the dataset includes $389,912$ papers, $851$ PACS numbers and $974,661$ classification labels.

The mathematics dataset is a collection of papers published from $1969$ to $2010$ and classified using the 2010 Mathematics Subject Classification. Here, we consider only those journal papers that have entries in both Mathematical Reviews and Web of Science. The MSC codes were obtained from the Mathematical Reviews records, and the other information was obtained from Web of Science. This dataset includes $705,574$ papers, $767$ MSC codes and $1,413,942$ classification labels.

The economics dataset is a collection of all economics papers collected by the American Economic Association Journal of Economic Literature from $1970$ to $2013$. Here, we consider only those papers that have records with both the JEL and Web of Science. The JEL Classification Codes were obtained from the Journal of Economic Literature, and the other information was obtained from Web of Science. This dataset includes $241,751$ papers, $129$ JEL codes and $411,865$ classification labels.

\paragraph{Scaling laws.} An allometric scaling-law relation between one quantity $Y_{s}$ and another quantity $N_{s}$ is assumed to have the form
\begin{equation}
\label{scaling}
Y_{s} = Y^{0}N_{s}^{\beta}
\end{equation}
$Y_{s}$ denotes an output (such as the number of papers) or input (such as the number of references) of a subfield $s$, and $N$ denotes the size of that subfield (such as the number of authors). $Y^{0}$ is a normalization constant. $\beta$ is the exponent, which we obtain through an ordinary least-squares (OLS) regression in log-log coordinates. The goodness of the regression is measured in terms of the coefficient of determination $R^2$, where $R$ is calculated as the correlation coefficient between $ln(Y_{s})$ and $ln(N_{s})$. This OLS analysis can be applied to the all-year data, in which case the values of $Y_{s}$ and $N_{s}$ are taken to be the cumulative values up through the last year covered by each dataset, or to single-year data. In the latter case, the exponent for year $t$ is denoted by $\beta\left(t\right)$.

Scaling laws, which follow a power-law function, describe the relation between two variables. In scaling laws,  the scaling exponents are generally obtained by OLS regression  \cite{Growth,Invention,interaction,road,Metropolitan,innovation1}. However, scaling laws are different from the recent interest in power laws, which generally describe probability distributions $P(x)\sim x^{-\alpha}$, e.g. the distribution of citations \cite{powerlaw,Innovation}. In power laws, due to the fact that to be a normalizable probability distribution function the power law usually holds only at the tail part of the distribution function and also due to noises in rare events at the very end of the tail part so that sometimes a cut-off has to be introduced, the exponents can be better estimated by the maximum likelihood method \cite{maxhood,maxhood1,maxhood2} than OLS regression. When there is a scaling law between two variables and when one of then two variables follows a power-law distribution, then clearly so does the other. Therefore, often scaling laws and power laws often appear together. However, this is not the case here in our analysis.

The relative stage of development of a subfield is measured in terms of the following deviation of the empirical value for that subfield with respect to the value predicted according to the allometric scaling relation:
\begin{equation}
\label{residuals}
\xi_{s} = ln\frac{Y_s}{Y^{0}N_s^{\beta}}
\end{equation}
It is independent of the absolute size of the subfield.

\paragraph{Author name disambiguation.} In the following, we investigate the possible scaling relationships between the number of papers and the numbers of author instances and authors, where the former simply counts the number of authors among all papers in a field regardless of whether some papers have the same or overlapping authors, whereas the latter counts only all unique authors. For the latter, we must address the problem of author disambiguation. In this paper, we adopt the simple \emph{last full and all initials} method to identify author names \cite{Authorname}, in which authors who have the same last name and all the same initials are considered to be the same author. For example, A Smith, AB Smith and AC Smith would be identified as distinct authors, but Alice Smith and Alysia Smith will be regarded as the same author.

The \emph{all initials} method has been claimed to have relatively low ``contamination" rates in certain disciplines, such as $1.5\%$ in mathematics and $2.2\%$ in economics \cite{Authorname}. We also performed our own small-scale validation of this approach. In the physics dataset, the subfield 42.50.Dv (Nonclassical states of the electromagnetic field) contains $13,005$ author-paper pairs. A total of $4,537$ distinct scientists were found after the disambiguation process.  To validate the \emph{all initials} method, we randomly selected 200 pairs of authors with
similar names, each consisting of two papers considered to be from the same author. We then verified whether they were indeed the same person by performing a search on the APS website and the authors' research homepages.  We found the false positive rate (i.e., the number of authors considered to be the same person whereas, in reality, they are not) to be $9\%$. We also performed a manual examination of the false negative rate (i.e., the number of identical authors incorrectly identified as different individuals using the all initials method) and found it to be approximately $4\%$.

\section{Results}

\begin{figure*}
\includegraphics[width=\linewidth]{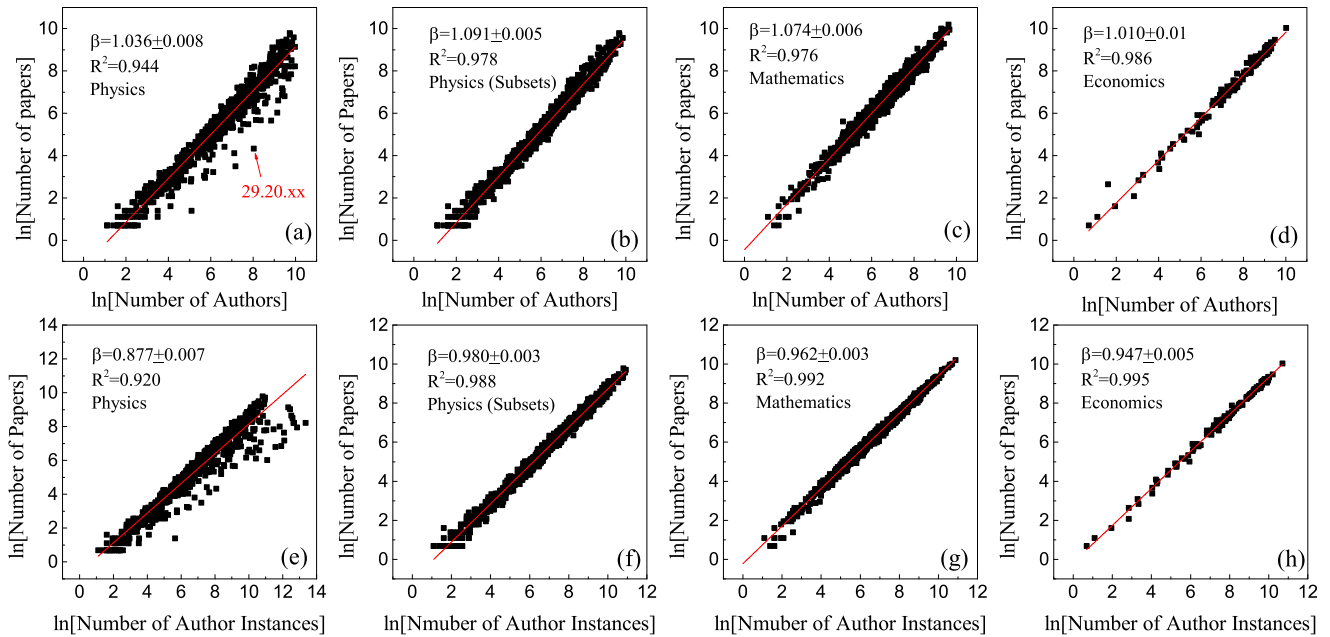}
\caption[\textbf{Scaling relations between the number of papers and the numbers of authors/author instances}]{\label{fig1} Scaling relations between the number of papers and the numbers of authors (author instances) for the subfields of physics (a) ((e)), physics (subsets) (b) ((f)), mathematics (c)((g)) and economics (d)((h)). $\ln{\left(x\right)}$ is the logarithm form of $x$. Each dot represents the accumulative numbers of publications and authors. }
\end{figure*}

First, let us consider the relation between the number of papers and the number of authors, which, in a sense, is similar to the relation between the output and size of cities. Here, we use the cumulative data up through the last year covered by the dataset for all three disciplines. We see that the values of the exponent $\beta$ are $1.036\pm0.008$ ($R^2=0.944$) for physics (Fig. \ref{fig1}(a)), $\beta=1.074\pm0.006$ ($R^2=0.976$) for mathematics (Fig. \ref{fig1}(c)) and $\beta=1.010\pm0.01$ ($R^2=0.986$) for economics  (Fig. \ref{fig1}(d)). This means that the number of papers per author very weakly increases as the number of authors increases. This, in turn, indicates that there are marginal increasing returns in physics, mathematics and economics.

When we consider Fig. \ref{fig1}(a) in further detail, we note that some subfields (29.20.xx) of physics are much less productive than predicted by the scaling law. For example, subfield 29.20.xx (\emph{Storage rings and colliders}) is related to high-energy experiments and has approximately $17.65$ authors per article on average. Such experimental subfields in physics generally require many scientists to work together. This might make the scaling exponent systemically smaller. To exclude these subfields, we restrict the analysis only to papers with at most ten authors (denoted by \emph{physics (subsets)}). As shown in Fig. \ref{fig1}(b), with this approach, the scaling exponent becomes $\beta=1.091\pm0.005$ ($R^2=0.978$). For cities, the scaling exponent $\beta$ between the number of new patents and the urban population is $1.27$, and that between the number of inventors and the urban population is $1.25$ \cite{Growth}. Therefore, we can roughly estimate the scaling exponent between the numbers of new patents and inventors to be approximately $1.27/1.25=1.016$. This rough estimation shows that our results are qualitatively consistent with the relation between the numbers of patents and inventors, which, in a sense, is similar to the relation between the numbers of papers and authors, as deduced from studies of scaling relations in cities \cite{Growth}. However, the exponent values of $1.03$ ($1.09$) in physics, $1.07$ in mathematics and $1.01$ in economics for the development of science/patents are quite different from the exponent relating the output and size of cities, which is roughly $\beta=1.25$. This means that the effect of increasing returns in science/patents is only marginal and not as high as the effect seen for production processes in cities. We do not know the reason for this difference. We can only speculate that it may be more difficult to increase scientific output than it is to increase industrial production by simply expanding in size.

\begin{figure*}
\includegraphics[width=0.5\linewidth]{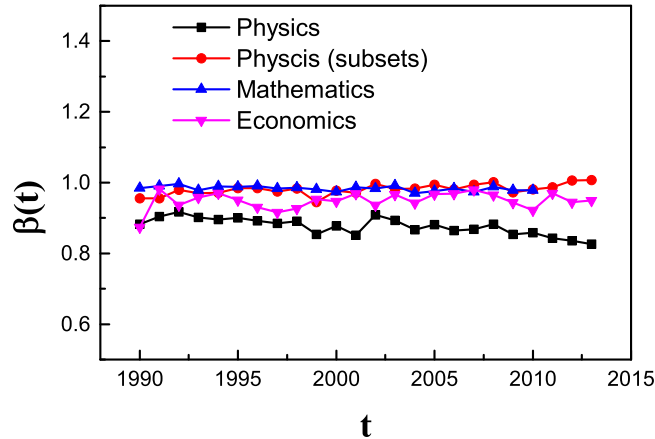}
\centering
\caption{\label{fig2} The evolution of the scaling exponents between the numbers of papers and authors. We see that except for physics, for which the value is smaller than the others and slightly decreasing, the exponents for the various fields have similar values that are stable over time. Note that the exponents from the yearly data, for some as-yet-unknown reason, are smaller than the exponents from the cumulative data.}
\end{figure*}

Next, let us check whether the scaling exponents have remained stable during all investigated years of development of the fields by performing a scaling analysis on the single-year data for each year. We know that the average numbers of authors and references in papers today are much larger than those in earlier times. However, we find that except for physics, for which the value is smaller than for the other fields and slightly decreasing, the values of the scaling exponent $\beta$ have remained quite stable, as shown in Fig. \ref{fig2}, especially for physics (subsets). The fact that similar scaling laws are observed in various disciplines implies that there might be a common mechanism governing the scientific progress of these disciplines, and the fact that the exponent values have remained similar and stable over time indicates that the underlying mechanism, if there is such a mechanism, tends to be preserved over time. The fact that physics as a whole shows a smaller and slightly decreasing exponent and the fact that we know that this phenomenon is due to papers with more than $10$ authors, which are often related to high-energy experimental physics, suggest that physics might have developed to a stage in which it often requires large teams to solve certain difficult problems and thus is less productive. It should be noted that the exponents in the yearly data analysis are smaller than the exponents for the cumulative data for reasons that we do not yet know.

Let us also compare this relation with the relation between the number of papers and the number of author instances (each appearance of an author, including duplicate authors, increments the total number of author instances by $1$). Interestingly, in this case, it is found that all
exponents are smaller than $1$, with $\beta= 0.877\pm0.007$ ($R^2=0.920$) for physics (Fig. \ref{fig1}(e)), $\beta= 0.980\pm0.003$ ($R^2=0.988$) for physics (subsets) (Fig. \ref{fig1}(f)), $\beta= 0.962\pm0.003$ ($R^2=0.992$) for mathematics (Fig. \ref{fig1}(g)) and $\beta= 0.947\pm0.005$ ($R^2=0.995$) for economics (Fig. \ref{fig1}(h)). This means that the marginal effect of increasing returns previously observed in Fig. \ref{fig1}(a-d) disappears when the number of author instances is considered, and the number of papers per author instance decreases as the number of author instances increases.

\begin{figure*}
\centering
\includegraphics[width=\linewidth]{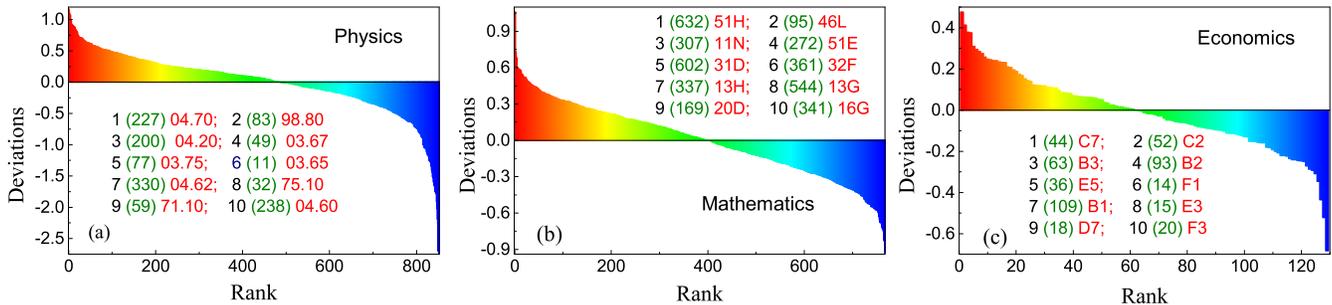}
\caption{\label{fig3} \textbf{Scale-independent rankings of subfields according to their deviations.} \textbf{(a) Physics} (1976-2013): the top $10$ subfields are 04.70 (Physics of black holes), 98.80 (Cosmology), 04.20 (Classical general relativity), 03.67 (Quantum information), 03.75 (Matter waves), 03.65 (Quantum mechanics), 04.62 (Quantum field theory in curved space time), 75.10 (General theory and models of magnetic ordering), 71.10 (Theories and models of many electron systems) and 04.60 (Quantum gravity). \textbf{(b) Mathematics} (1969-2010): the top $10$ subfields are 51H (Topological geometry), 46L (Self-adjoint operator algebras), 11N (Multiplicative number theory), 51E (Finite geometry and special incidence structures), 31D (Axiomatic potential theory), 32F (Geometric convexity), 13H (Local rings and semilocal rings), 13G (Integral domains), 20D (Abstract finite groups), and 16G (Representation theory of rings and algebras). \textbf{(c) Economics} (1970-2013): the top $10$ subfields are C7 (Game Theory and Bargaining Theory), C2 (Single Equation Models; Single Variables), B3 (History of Thought: Individuals), B2 (History of Economic Thought since 1925), E5 (Monetary Policy, Central Banking, and the Supply of Money), F1 (Trade (International Economics), B1 (History of Economic Thought through 1925), E3 (Prices, Business Fluctuations, and Cycles), D7 (Analysis of Collective Decision-Making), and F3 (International Finance). The numbers in brackets are the ranks of the various subfields according to their sizes.}
\end{figure*}

Although the goodness of fit of the fitted curves are very high overall, there are some outliers that are relatively far from the fitted curves in the above figures, and the relative positions of the subfields often change from year to year. The residual is a measure of the deviation of a true value from the corresponding value predicted by the scaling law (Eq. (\ref{residuals})). These deviations provide a meaningful way to rank cities \cite{interaction} and universities \cite{university2}. In Fig. \ref{fig3}, we show the ranking of the deviations by magnitude and sign for physics and economics in 2013 as well as those for mathematics in 2010. Let us focus on a few subfields that deviate strongly and positively from the scaling law. For example, the output of classical general relativity (04.20) is ranked 3rd in physics in Fig. \ref{fig3}(a) according to its deviation, but it is a relatively small subfield (ranked 200th by size). Quantum information (03.67) ranks 4th in physics in Fig. \ref{fig3}(a) but is ranked 49th according to its size. When they are ranked according to their sizes, classical general relativity is not ranked similarly to quantum information. However, when they are ranked according to their deviations, we see that they are both among the top $10$ subfields in physics. These findings are broadly consistent with our intuition regarding these subfields: one is small and one is big, but both are very active subfields. This implies that, at least in part, the deviation from the fitted scaling law provides a reasonable indicator of the ranking of the subfields that is independent of their sizes.

We also rank the subfields of mathematics and economics. It is found that topological geometry (51H) is ranked 1st in mathematics according to its deviation, whereas it is ranked 632nd according to its size. In addition, it is found that game theory and bargaining theory (C7) is the top subfield in economics according to its deviation but is ranked 44th according to its size. Judging from our limited knowledge of economics, we believe that it is reasonable for game theory to be considered among the top subfields: it is not large but is a core subfield of economics, which can partially be seen from the fact that (according to Wikipedia) there have been $11$ game theorists among the Nobel laureates in economics, and some economists even believe that it is the core of the whole of economic theory \cite{Levin:GameTheory}.

\begin{figure*}
\includegraphics[width=\linewidth]{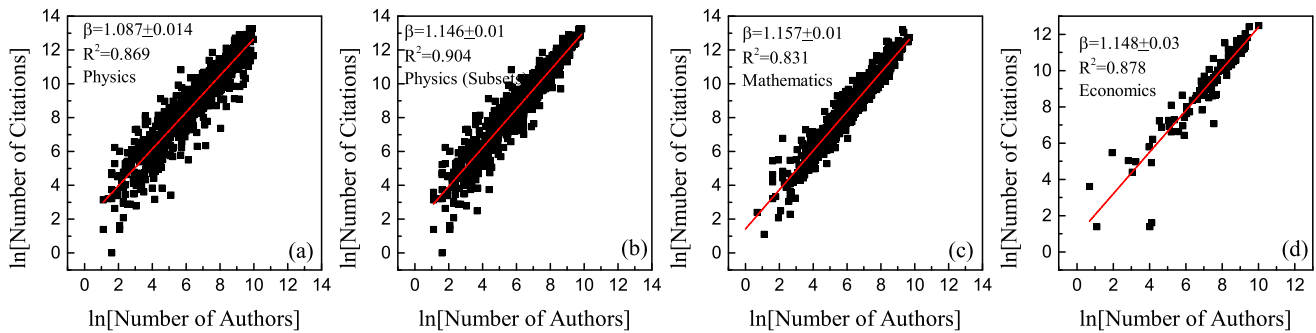}
\caption{\label{fig4} Scaling relations between the numbers of citations and authors in physics (a), physics (subsets) (b), mathematics (c) and economics (d).  }
\end{figure*}

Let us now look at other outputs of the investigated scientific fields vs. the numbers of authors in their subfields. It is found that the exponents relating the numbers of citations and authors are larger than $1$, with $\beta = 1.087\pm0.014$ ($R^2=0.869$) for physics, $\beta = 1.146\pm0.01$ ($R^2=0.904$) for physics (subsets), $\beta = 1.157\pm0.01$ ($R^2=0.831$) for mathematics and $\beta = 1.148\pm0.03$ ($R^2=0.878$) for economics (Fig. \ref{fig4}). This means that authors working in larger subfields receive, on average, more citations than those in smaller subfields. In addition, the exponents relating the numbers of citations and papers are $\beta = 1.059\pm0.009$ ($R^2=0.934$) for physics, $\beta = 1.063\pm0.009$ ($R^2=0.935$) for physics (subsets), $\beta = 1.059\pm0.011$ ($R^2=0.920$) for mathematics and $\beta = 1.117\pm0.036$ ($R^2=0.939$) for economics. These findings are similar to the scaling laws between the numbers of citations and papers when universities \cite{university1,university2} and research groups \cite{group1,group2,group3} are treated as the relevant units. However, the exponent values for the latter cases are approximately $\beta \approx 1.25$, larger than those found here. This means that whereas authors are more likely to cite papers from the same university, the same research group and the same subfield, the degrees of affinity for universities and research groups are even stronger than those for subfields.

\begin{figure*}
\includegraphics[width=\linewidth]{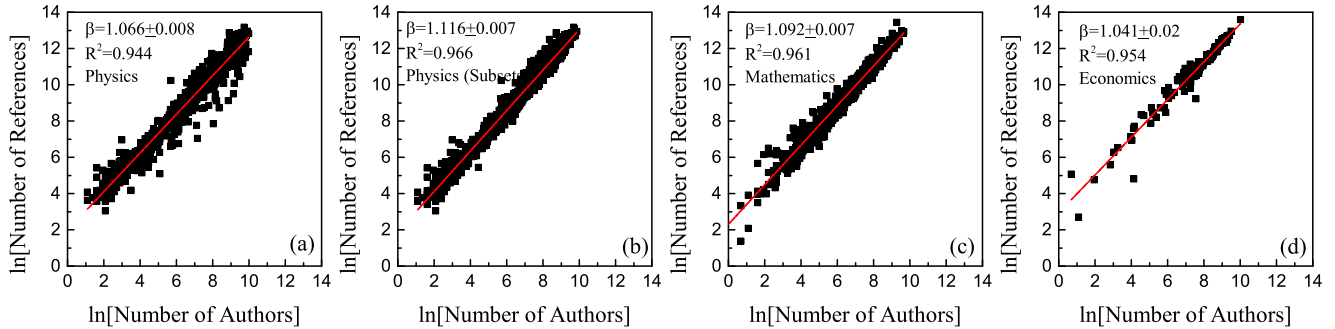}
\caption{\label{fig5} Scaling relations between the numbers of references and authors in physics (a), physics (subsets) (b), mathematics (c) and economics (d). }
\end{figure*}

Next, we find that the scaling-law exponent relating the numbers of references and authors is smaller than the exponent between numbers of citations and authors. We have $\beta=1.066\pm0.008$ ($R^2=0.944$) for physics, $\beta=1.116\pm0.007$ ($R^2=0.966$) for physics (subsets), $\beta=1.092\pm0.007$ ($R^2=0.961$) for mathematics and $\beta=1.041\pm0.02$ ($R^2=0.954$) for economics (Fig. \ref{fig5}). In scaling law of cities, similarly the exponent of supplies and population is also smaller than the exponent of outputs and population\cite{Growth}. We might expect these exponents in the case of scientific publications to be higher since, intentionally or unintentionally, people may cite references more carelessly than they would use living supplies because there is no cost for citing more references, whereas there is a cost associated with the use of living supplies. However, the fact that these exponents are close to, although clearly slightly higher than, that relating the housing/water/energy supplies and populations in cities implies that perhaps researchers do not cite many unnecessary references.

\section{Conclusions and Discussion}

In this paper, we first examined and confirmed the allometric scaling relations between the numbers of papers, citations, and references and numbers of authors in subfields of three disciplines, namely, physics, mathematics and economics, which are analogous to the relation between the numbers of patents and inventors of patents in cities \cite{Growth} and the relations between various outputs/inputs and population size for cities \cite{Growth} and countries \cite{countries}. One of the reasons for the development of cities is that there is an effect of increasing returns between the output and size of a city, which results in a lower effective cost for intra-city transactions than for inter-city transactions. Perhaps there are similar factors driving the formation of scientific subfields, which cause the observed allometric scaling relations to arise in the development of research subfields. Furthermore, the values of the exponents for all three disciplines were found to be similar and to have remained stable over time. We do not yet know why the various disciplines display similar exponents and temporal stability. However, we believe that this common allometric law across disciplines and time requires further investigation: certain common underlying mechanisms may exist that drive the development of scientific fields in various disciplines.

We found that the exponents relating the numbers of papers and authors are much smaller than those relating the various outputs of cities to their size \cite{Growth}. This means that the effect of increasing returns observed in scientific production is much lower than the corresponding effect on production in cities. However, the exponents relating the numbers of citations and authors are more similar to those relating the various outputs of cities to their size, indicating that there is a stronger effect of increasing returns between the numbers of citations and authors. This suggests that on average, as the number of authors increases, there is only a marginal effect of increasing returns on the number of papers but a much larger effect of increasing returns on the number of citations. In addition, through several examples, we showed that deviations of individual subfields from the predictions of the allometric scaling relations can provide a size-independent but still meaningful ranking of those subfields.

The current study has several limitations. Our datasets contained only the portions of WOS that overlap with the relevant subject classification schemes (PACS, MSC and JEL), which restricted our ability to study the scaling relations governing the properties of all publications. In particular, the results for physics consider only those papers published in Physical Review journals. Moreover, the method used for author name disambiguation could be further improved.

\end{document}